\documentclass[floatfix,aps,prd,showpacs,amsmath,nofootinbib,twocolumn]{revtex4}
\usepackage{epsfig}
\newcommand{\be}{\begin{equation}}
\newcommand{\ee}{\end{equation}}

\begin{document}

\title{Fourth order Weyl Gravity}
\author{\'Eanna \'E. Flanagan}\email{eef3@cornell.edu}
\affiliation{
Laboratory for Elementary Particle Physics, Cornell University, Ithaca, NY 14853, USA.}

\date{\today}

\begin{abstract}
The fourth order Weyl gravity theory of Mannheim and Kazanas is based
on replacing the Einstein-Hilbert action with the square of the Weyl
tensor, and on modifying the matter action of the standard model of
particle physics to make it conformally invariant.  This theory has
been suggested as a model of both dark matter and dark energy.  We
argue that the conformal invariance is not a fundamental property of
the theory, and instead is an artifact of the choice of variables used
in its description.  We deduce that in the limit of weak
fields and slow motions the theory does not agree with
the predictions of general relativity, and is therefore ruled out by
Solar System observations.

\end{abstract}

\pacs{95.36.+x,04.50.+h,95.35.+d}

\maketitle

\def\be{\begin{equation}}
\def\ee{\end{equation}}
\def\bes{\begin{subequations}}
\def\ees{\end{subequations}}
\def\bea{\begin{eqnarray}}
\def\eea{\end{eqnarray}}

\def\nn{\nonumber}
\maketitle



\def\alt{
\mathrel{\raise.3ex\hbox{$<$}\mkern-14mu\lower0.6ex\hbox{$\sim$}}
}

\section{Introduction}

\label{sec1}
In the fourth order Weyl theory of gravity
\cite{weyl1,weyl2,weyl3,weyl4,weyl5,weyl6,weyl7,weyl8,review}, the
Einstein-Hilbert action is replaced by a term proportional to the
square of the Weyl tensor, and the action of the standard model of
particle physics is modified to make it be conformally invariant.
For simplicity and following Ref.\ \cite{review} we work here with a
subset of the standard model consisting of a Dirac fermion field
$\psi$, a gauge field $A_\alpha$ and a real scalar field $S$ which
plays the role of the Higgs field.  The action of the theory is a
functional of these fields and of a metric $g_{\alpha\beta}$:
\begin{eqnarray}
&& S[g_{\alpha\beta},S,\psi,A_\alpha] = \int d^4 x \sqrt{-g} \bigg\{
  -\alpha_g C_{\alpha\beta\gamma\delta} C^{\alpha\beta\gamma\delta}
  \nn \\
&&- {1 \over 2} \nabla_\alpha S \nabla^\alpha S - {1 \over 12} S^2 R -
\lambda S^4  - {1 \over 4} F_{\mu\nu} F^{\mu\nu} \nonumber \\
&&  + {i \over 2}  {\bar \psi} \gamma^\mu \nabla_\mu \psi
- {i \over 2} {\overline {\nabla_\mu \psi}} \gamma^\mu \psi
+ e {\bar \psi} \gamma^\mu A_\mu \psi
- h S {\bar \psi} \psi \bigg\}. \nn \\
\mbox{}
\label{action1}
\end{eqnarray}
Here $\alpha_g$, $h$ and $\lambda$ are dimensionless parameters, we
use natural units with $\hbar = c =1$, and we use the sign convention
$(+,+,+)$ in the notation of Ref.\ \cite{MTW} \footnote{Our sign
  conventions differ from those used by Mannheim in Ref.\
  \cite{review}}.

The motivations for the action (\ref{action1}) are as follows \cite{review}.  First, it
is invariant under the conformal transformations
\bes
\bea
g_{\alpha\beta} &\to& e^{2 \sigma} g_{\alpha\beta}  \\
S &\to& e^{-\sigma} S  \\
\psi &\to& e^{-3 \sigma/2} \psi \\
A_\alpha &\to& A_\alpha,
\eea
\label{symmetry}
\ees
where $\sigma = \sigma(x)$ is arbitrary. This exact symmetry prevents
the appearance of a cosmological constant.  Second, it was argued that
the term coupling the Ricci scalar to the scalar field $S$ can drive a
gravity-mediated spontaneous symmetry breaking:
namely, in
the presence of a background value of $R$, the minimum energy state of
$S$ will occur at a nonzero value of $S$ and will thereby give mass to
the fermion field.  In Refs.\ \cite{weyl1,weyl2,weyl5,weyl6} it was
argued that the theory (\ref{action1}) agrees with observations of
Newtonian gravity in the Solar System, and in addition predicts a
linearly growing term in the Newtonian potential that could explain
galactic rotation curves without the need for dark matter.  Refs.\
\cite{weyl7,weyl8,review} argue that Weyl gravity yields a viable
model of the acceleration of the Universe, removing the need for dark
energy.  Further studies of the theory can be found in Refs.\
\cite{study1,study2,study3,study4,study5,study6,study6a,study7,study8,study9,study10}.

In this paper we rewrite the theory (\ref{action1}) in a new set of
variables that allows a simpler computation of its predictions.  We
also show that the theory does not reproduce Solar System observations in the
limit of weak fields and slow motions, which rules out the theory.

\section{Reformulation of theory}

We specialize at first to the sector of the theory where
\be
S(x) >0
\label{sector1}
\ee
everywhere \footnote{It follows from our analysis below that if $S$ is
positive on an initial data surface, then it is positive throughout
spacetime.}.  We define the new variables
\begin{subequations}
\begin{eqnarray}
{\hat g}_{\alpha\beta} &=& {S^2 \over m_0^2} g_{\alpha\beta} \\
{\hat S} &=&  S \\
{\hat \psi} &=& {m_0^{3/2} \over S^{3/2}} \psi \\
{\hat A}_\alpha &=& A_\alpha,
\end{eqnarray}
\end{subequations}
where $m_0$ is an arbitrary but fixed positive parameter with dimensions of mass.
All of these variables, except ${\hat S}$, are conformal invariants.  The
action in terms of the new variables is\footnote{Here ${\hat
    \gamma}^\mu$ are the Dirac matrices associated with the
metric ${\hat g}_{\mu\nu}$ that satisfy $\{ {\hat \gamma}^\mu,{\hat
  \gamma}^\nu \} = - 2 {\hat g}^{\mu\nu}$.}
\begin{eqnarray}
&& S[{\hat g}_{\alpha\beta},{\hat S},{\hat \psi},{\hat A}_\alpha] =
\int d^4 x \sqrt{-{\hat g}} \bigg\{
  -\alpha_g {\hat C}_{\alpha\beta\gamma\delta} {\hat C}^{\alpha\beta\gamma\delta}
  \nn \\
&&- {1 \over 12} m_0^2 {\hat R} -
\lambda m_0^4  - {1 \over 4} {\hat F}_{\mu\nu} {\hat F}^{\mu\nu} \nonumber \\
&&
 + {i \over 2}  {\bar {\hat \psi}} {\hat \gamma}^\mu {\hat
\nabla}_\mu {\hat \psi}
- {i \over 2} {\overline {{\hat \nabla}_\mu {\hat \psi}}} {\hat
  \gamma}^\mu {\hat \psi}
+ e {\bar {\hat \psi}} {\hat \gamma}^\mu {\hat A}_\mu {\hat \psi}
- h m_0 {\bar {\hat \psi}} {\hat \psi}
\bigg\}. \nn \\
\mbox{}
\label{action1a}
\end{eqnarray}

In this new representation, the only field which transforms under the
conformal symmetry (\ref{symmetry}) is ${\hat S}$.  However, the
action (\ref{action1a}) is independent of ${\hat S}$.  Thus there are
two uncoupled sectors of the theory (\ref{action1}) with the
constraint (\ref{sector1}):
a trivial sector containing
${\hat S}$ on which the symmetry acts, and which contains no dynamics;
and the remaining sector containing the fields ${\hat
  g}_{\alpha\beta}$, ${\hat A}_\alpha$ and ${\hat \psi}$, which does
not possess a conformal symmetry.
For the remainder of this paper, we will drop the field ${\hat S}$ and
consider only the dynamical sector of the theory.

In a similar manner, one can start from the action for general relativity coupled
to the standard model of particle physics, perform the above
operations in reverse, and obtain an equivalent action with one extra
scalar field which has an exact conformal symmetry.
It follows
that the conformal symmetry of the theory (\ref{action1})
is not a fundamental or defining property
of the theory, and is instead an artifact of the
choice of variables used
to describe the theory.

The transition from the action (\ref{action1}) to the action
(\ref{action1a}) can also be thought of as a gauge fixing
\cite{barabash,wood}.  The conformal symmetry (\ref{symmetry}) is
analogous to a gauge freedom, and we are free to analyze the theory in
the gauge $S(x) = m_0$, which leads to the action (\ref{action1a}).

In the action (\ref{action1a}), the parameter $m_0$ can be chosen
arbitrarily.  This arbitrariness is the freedom of choice of
units of mass.  Only the ratios of the three mass parameters which
appear in the action are measurable.  These three mass parameters are
the Planck mass (the coefficient of Ricci scalar), the
cosmological constant term, and the mass term for the fermion field.
If we define $m_p^2  =  m_0^2/6$, $\Lambda = \lambda m_0^4$, and $m_e
= h m_0$, then the action can be written in the more familiar looking form
\begin{eqnarray}
&& S[{\hat g}_{\alpha\beta},{\hat S},{\hat \psi},{\hat A}_\alpha] =
\int d^4 x \sqrt{-{\hat g}} \bigg\{
  -\alpha_g {\hat C}_{\alpha\beta\gamma\delta} {\hat C}^{\alpha\beta\gamma\delta}
  \nn \\
&&- {1 \over 2} m_p^2 {\hat R} -
\Lambda  - {1 \over 4} {\hat F}_{\mu\nu} {\hat F}^{\mu\nu} \nonumber \\
&&
 + {i \over 2}  {\bar {\hat \psi}} {\hat \gamma}^\mu {\hat
\nabla}_\mu {\hat \psi}
- {i \over 2} {\overline {{\hat \nabla}_\mu {\hat \psi}}} {\hat
  \gamma}^\mu {\hat \psi}
+ e {\bar {\hat \psi}} {\hat \gamma}^\mu {\hat A}_\mu {\hat \psi}
-  m_e {\bar {\hat \psi}} {\hat \psi}
\bigg\}. \nn \\
\mbox{}
\label{action2}
\end{eqnarray}
This is the standard action for a fermion field of mass $m_e$ and charge
$e$ coupled to a gauge field and coupled to general
relativity, except for three modifications to the gravitational part
of the action: (i) the addition of the cosmological constant term; (ii)
the sign of the Ricci term is flipped, and (iii) the Weyl squared term
is added.  Note also that the value of the Planck mass parameter $m_p$
can be different from its conventional value of $\sim 10^{19}$ GeV;
below we will consider all possible values of the parameters
$\alpha_g$ and $m_p$.

We next return to the original action (\ref{action1}), and consider
the sector of the theory where $S(x) < 0$ everywhere.  A similar
analysis shows that this sector is also described by an action of the form
(\ref{action2}), but with the sign of the fermion mass term flipped.
This can be compensated for by redefining $\psi \to \gamma^5 \psi$.
Thus the $S<0$ sector behaves the same way as the $S>0$ sector.  We
will confine attention to the $S>0$ sector.

\section{Weak field limit}

Consider now the predictions of the theory (\ref{action2}) in the
limit of weak fields
and slow motions.  A key point is that the physical metric measured by
experiments\footnote{Here it is assumed that the units for length and
time are defined using non-gravitational physics.  Systems of units
in common use such as SI units satisfy this requirement.  } is the metric
${\hat g}_{\alpha\beta}$, and not the metric $g_{\alpha\beta}$ that
appeared in the original action (\ref{action1}).  This follows from
the form of the action (\ref{action2}), which has a standard form that
implies that objects constructed from the fermion and gauge fields will
fall on geodesics of ${\hat g}_{\alpha\beta}$ \footnote{It is
immediately clear that $g_{\alpha\beta}$ cannot be the physical
metric, since the theory cannot predict $g_{\alpha\beta}$ uniquely,
only $g_{\alpha\beta}$ up to conformal transformations.  By contrast,
the metric ${\hat g}_{\alpha\beta}$ can be predicted uniquely.}.
Equivalently, in terms of the original variables $g_{\alpha\beta}$ and $S$,
all freely falling objects are subject to an acceleration proportional
to the gradient of $S$, as argued by Wood \cite{wood}.

It is straightforward to show that the theory (\ref{action2}) does not
admit a regime in which its predictions agree with the weak-field slow-motion limit of
general relativity, for any choice of values of the
parameters $\alpha_g$ and $m_p$, which implies that the theory is
ruled out by Solar System
observations.  Substituting the ansatz
\bes
\bea
\label{metricansatz}
{\hat g}_{ab} dx^a dx^b &=& - [1 + 2 \Phi({\bf x})] dt^2 + [1 - 2
  \Psi({\bf x}) ] \delta_{ij} dx^i dx^j \nn \\
&& \\
T_{ab} dx^a dx^b &=& \rho({\bf x}) dt^2
\eea
\ees
into the linearized equation of motion obtained from the action
(\ref{action2}) yields the solution
\bes
\bea
\label{Phisoln}
\Phi &=& {4 \over 3} {\bar \Phi} + {1 \over 3} \Phi_{\rm N} \\
\Psi &=& {2 \over 3} {\bar \Phi} - {1 \over 3} \Phi_{\rm N}.
\eea
\label{soln1}
\ees
Here $\Phi_{\rm N}$ is the usual Newtonian potential which satisfies
\be
2 m_p^2 \nabla^2 \Phi_{\rm N} = \rho,
\label{PhiNdef}
\ee
where $\rho$ is the mass density, and ${\bar \Phi}$ is a potential
which satisfies the fourth order equation
\be
8 \alpha_g \nabla^2 \nabla^2 {\bar \Phi} - 2 m_p^2 \nabla^2 {\bar
  \Phi} = \rho.
\label{eom}
\ee
We
have neglected the cosmological constant term whose influence will be
negligible on Solar System scales and smaller scales.

\begin{figure}
\begin{center}
\epsfig{file=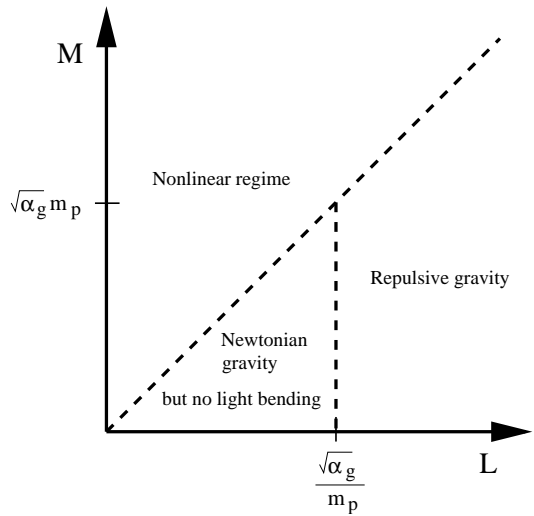,width=7cm}
\caption{The various regimes for fourth order Weyl gravity for a
  source of mass $\sim M$ and size $\sim L$.  }
\label{figure1}
\end{center}
\end{figure}

We now consider an isolated source of mass $\sim M$ and size $\sim
L$.  Some useful information can be obtained from dimensional analysis.
In a general system of units with $\hbar \ne 1$, the action
(\ref{action2}) contains two independent dimensionful parameters, the mass scale
$\sqrt{\alpha_g} m_p$ and the lengthscale $\sqrt{\alpha_g} / m_p$.
There are therefore two dimensionless parameters characterizing the
source, namely $L m_p / \sqrt{\alpha_g}$ and $M / (\sqrt{\alpha_g}
m_p)$.  There are three different regimes in this two-dimensional
parameter space in which the theory exhibits different types of
behavior (see Fig.\ \ref{figure1}):
\bes
\bea
\label{regimeI}
{\rm regime\ 1:}&\ \ \ {L m_p \over \sqrt{\alpha_g}} \gg {M \over \sqrt{\alpha_g} m_p}, \ \ \  &
{L m_p \over \sqrt{\alpha_g}}  \gg 1 \\
\label{regimeII}
{\rm regime\ 2:}&\ \ \ {L m_p \over \sqrt{\alpha_g}} \gg {M \over \sqrt{\alpha_g} m_p}, \ \ \  &
{L m_p \over \sqrt{\alpha_g}}  \ll 1 \\
\label{regimeIII}
{\rm regime\ 3:}&\ \ \ {L m_p \over \sqrt{\alpha_g}} \alt {M \over \sqrt{\alpha_g} m_p}. \ \ \  &
\eea
\label{regimes}
\ees
We now discuss these various regimes in turn.

\subsection{Regime 1}

Consider the ratio between the first and
second terms in Eq.\ (\ref{eom}), evaluated in the vicinity of the
source at $r \sim L$.  This ratio is $\sim \alpha_g / ( m_p^2 L^2)$,
which is small compared to unity by Eq.\ (\ref{regimeI}).  Consequently
the fourth order derivative term gives only a
small correction, and it follows that ${\bar \Phi} \approx - \Phi_{\rm N}$ and
so also $\Phi = \Psi = - \Phi_{\rm N}$ from Eqs.\ (\ref{soln1}).
Thus Newton's law of
gravitation is recovered but with the sign flipped.  The resulting repulsive
gravitational force disagrees with observations.  This conclusion applies
in particular for the conventional values of the parameters, namely $\alpha_g \sim
1$ and $m_p \sim 10^{19}$ GeV.

\subsection{Regime 2}

The exact solution to Eq.\ (\ref{eom}) is
\be
{\bar \Phi}({\bf x}) = - {1 \over 4 \pi} \int d^3 y { \chi({\bf y}) \over |
  {\bf x} - {\bf y}|},
\label{phisoln}
\ee
where\footnote{If $\alpha_g$ is negative rather than positive as
  assumed here, the exponential in Eq.\ (\protect{\ref{psi}}) is
  replaced by a cosine.  This replacement does not qualitatively
  change the subsequent discussion.}
\be
\chi({\bf y}) = - {1 \over 32 \pi \alpha_g} \int d^3z { e^{- {1
      \over \sqrt{2} \alpha_g} m_p | {\bf y} - {\bf z} | } \over | {\bf y} -
  {\bf z} | } \rho({\bf z}).
\label{psi}
\ee
Using this solution we obtain the order of magnitude estimates
$\chi(r) \sim M / (\alpha_g L)$ for $r \sim L$, $\chi(r) \sim M /
(\alpha_g r)$ for $L \ll r \ll \sqrt{\alpha_g} /m_p$, while $\chi$
falls off exponentially for $r \agt \sqrt{\alpha_g} /m_p$.
In spherical symmetry the gradient of the field ${\bar \Phi}$ is
\be
{\partial {\bar \Phi} \over \partial r } \sim {1 \over r^2} \int_0^r
dr' (r')^2 \psi(r')
\label{acc}
\ee
which yields the estimate ${\bar \Phi}_{,r} \sim M / \alpha_g$ for $r
\sim L$, ${\bar \Phi}_{,r} \sim M / \alpha_g$ for $L \ll r \ll
\sqrt{\alpha_g} / m_p$, and ${\bar \Phi}_{,r} \sim M / (m_p^2 r^2)$ for
$r \gg \sqrt{\alpha_g}/m_p$.  Therefore for $r \ll
\sqrt{\alpha_g}/m_p$ the acceleration produced by
the potential ${\bar \Phi}$ is smaller than the acceleration $\sim M /
(m_p^2 r^2)$ produced by the Newtonian potential term $\Phi_{\rm N}$
in the expression (\ref{Phisoln}) for $\Phi$ by a factor of $\sim
m_p^2 r^2 / \alpha_g \ll 1$.  Therefore to a good approximation the
solution in this regime is given by\footnote{Up to an overall constant
term in ${\bar \Phi}$ which can be eliminated
by a gauge transformation of the form $x^i \to \alpha x^i$, $t \to
\alpha^{-2} t$ and which is not locally measurable.} Eqs.\ (\ref{soln1}) with the
${\bar \Phi}$ terms dropped:
\be
\Phi = {1 \over 3} \Phi_{\rm N}, \ \ \ \ \Psi = - {1 \over 3}
\Phi_{\rm N}.
\label{soln4}
\ee
Since the motion of massive particles is governed by the potential
  $\Phi$, we see that Newtonian gravity is recovered locally for
  massive particles with an effective Newton's constant $G_{\rm eff} =
  1 / (24 \pi m_p^2)$.

The problem which occurs in regime 2 is light bending.  Since the
metric given by Eqs.\ (\ref{metricansatz}) and (\ref{soln4}) is
conformally flat to a good approximation, there is no light bending.
More precisely, the amount of light bending produced for a ray that
grazes the source is smaller than the prediction of general relativity
by a factor of $\sim L^2 m_p^2 / \alpha_g \ll 1$.
Another way of describing this is in terms of the Eddington PPN
parameter $\gamma$, defined by the metric expansion in spherical symmetry
\bea
ds^2 &=& - \left[1 - {2 {\bar M} \over r} + O(1 / r^2) \right] dt^2 \nn\\
&&+ \left[ 1
  + {2 \gamma {\bar M} \over r} + O(1/r^2) \right] \delta_{ij} dx^i dx^j.
\label{gammadef}
\eea
Comparing Eqs.\ (\ref{metricansatz}) and (\ref{soln4}) with Eq.\ (\ref{gammadef})
yields
\be
\gamma = -1 + O \left( {L^2 m_p^2 \over \alpha_g} \right).
\ee
Experimentally it is known that $\gamma=1$ to a within a small
fraction of a percent, in agreement with the prediction of general relativity.
The deflection of a ray of light is proportional to $1 + \gamma$.

\subsection{Regime 3}

We first note that the linearized equations of motion
are a good approximation in regimes 1 and 2.  The potential $\Phi_{\rm
  N} \sim M / (m_p^2 L)$ defined by Eq.\ (\ref{PhiNdef}) is small
compared to unity by
Eqs.\ (\ref{regimeI}) and (\ref{regimeII}).  In regime 1 we have
${\bar \Phi} \approx - \Phi_{\rm N}$ so $|{\bar \Phi}| \ll 1$.  In
regime 2, the largest value of ${\bar \Phi}$ is of order $(M/\alpha_g)
(\sqrt{\alpha_g}/m_p) \sim M \sqrt{\alpha_g} / m_p$ which is small
compared to unity by Eqs.\ (\ref{regimeII}).

In regime 3 however, we have $|\Phi_N| \agt 1$ from Eq.\
(\ref{regimeIII}), and so the linearized approximation breaks down.
In this regime one must use the full nonlinear equations of the theory.
However, it is clear  that Newtonian phenomenology cannot be
reproduced in this regime since the linear superposition principle
will not apply.

\medskip

Finally, a separate problem with the theory (\ref{action2}) is that it
contains a ghost field, i.e. a field whose
kinetic energy term has the wrong sign.  This ghost field is a
massless spin 2 field that is due to the negative Ricci term
in the action (\ref{action2}) \cite{barabash}.  There is also a
massive spin 2 field in the theory;
this field is normally ghostlike \cite{chiba} but here is not, due to the negative
coefficient of the Ricci scalar.  It is however tachyonic for $\alpha_g>0$.

As this paper was being completed, we learned that similar arguments
had been presented by Karel Van Acolyen at a conference
\cite{workshop}.

\acknowledgements

We thank Philip Mannheim and Karel Van Acoleyen
for helpful conversations.
This research was sponsored in part by NSF grant PHY-0457200.

\end{document}